\documentclass[12pt]{iopart}
\usepackage{graphicx}
\expandafter\let\csname equation*\endcsname=\relax 
\expandafter\let\csname endequation*\endcsname=\relax 
\usepackage{amsmath}
\usepackage[colorlinks]{hyperref}
\usepackage{cleveref}
\usepackage{cancel}
\usepackage[normalem]{ulem}
\usepackage{xcolor}

\newcommand{\ffrac}[2]{\ensuremath{\frac{\displaystyle #1}{\displaystyle #2}}}
\newcommand{\blb}{\ensuremath{\bar\lambda}}
\newcommand{\bmu}{\ensuremath{\bar\mu}}
\newcommand{\blbc}{\ensuremath\blb_\mathrm{c}}
\newcommand{\av}[1]{\ensuremath\langle #1\rangle}
\newcommand{\kmax}{k_\textrm{max}}
\newcommand{\derivative}[2]{\ffrac{\rmd#1}{\rmd#2}}
\newcommand{\nni}[1]{n_{#1}}
\newcommand{\nnik}[2]{n_{#1}^{[#2]}}
\newcommand{\Iik}[2]{I_{#1}^{[#2]}}
\newcommand{\Iikeff}[2]{\tilde{I}_{#1}^{[#2]}}
\newcommand{\nnieff}[1]{\tilde n_{#1}}
\newcommand{\nnikeff}[2]{\tilde n_{#1}^{[#2]}}

\newcommand{\rinfk}[2]{\rho_{#1}^{[#2]}}

\newcommand{\rinfkeff}[2]{\tilde \rho_{#1}^{[#2]}}
\newcommand{\pok}[1]{P_{#1}}
\newcommand{\pokeff}[1]{\tilde P_{#1}}

\newcommand{\knieff}[2]{\widetilde{\av{k^{#1}}}_{#2}}
\newcommand{\rinfkst}[2]{{\rho_{#1}^{*}}^{[#2]}}

\newcommand{\pik}[2]{\Pi_{#1}^{[#2]}}
\newcommand{\pikst}[2]{{\Pi^{*}_{#1}}^{[#2]}}
\newcommand{\Qik}[2]{\pi_{#1}^{[#2]}}
\newcommand{\Qikst}[2]{{\pi^{*}_{#1}}^{[#2]}}
\newcommand{\rinfkeffst}[2]{{\tilde{\rho}_{#1}}^{*[#2]}}

\newcommand{\epik}[2]{\bar{\epsilon}_{#1#2}}
\newcommand{\bepik}[1]{{\epsilon}_{#1}}
\newcommand{\Mijkkl}{M_{ik}^{jk'}}
\newcommand{\Mij}{M_{ij}}
\newcommand{\tMij}[1]{\widetilde{M_{ij}^{(#1)}}}
\newcommand{\tM}[1]{\widetilde{{\textbf{M}}^{(#1)}}}
\newcommand{\bMijkkl}{\bar{M}_{ik}^{jk'}}

\newcommand{\nkm}[1]{n_{#1}\av{k}_{#1}}

\newcommand{\Lb}[1]{\Lambda^{(#1)}}

\newcommand{\dij}{\delta_{ij}}

\begin{document}
	
	\title[Disease dynamics with heterogeneous transmission and recurrent mobility patterns]{Infectious disease dynamics in metapopulations with heterogeneous transmission and recurrent mobility}
	
	\author{Wesley Cota$^{1,2}$, David Soriano-Pa\~nos$^{2,3}$, A. Arenas$^{4}$, Silvio C. Ferreira$^{1,5}$, Jes\'us G\'omez-Garde\~nes$^{2,3,6}$}
	
	\address{$^1$ Departamento de Fisica, Universidade Federal de Vi\c{c}osa, 36570-900 Vi\c{c}osa, Minas Gerais, Brazil}
	\address{$^2$ GOTHAM Lab -- Institute of Biocomputation and Physics of Complex Systems (BIFI), University of Zaragoza, E-50018 Zaragoza, Spain}
	\address{$^3$ Department of Condensed Matter Physics, University of Zaragoza, E-50009 Zaragoza, Spain}
	\address{$^4$ Department of Computer Science and Mathematics, Universitat Rovira i Virgili, 43007 Tarragona, Spain}
	\address{$^5$ National Institute of Science and Technology for Complex Systems, 22290-180 Rio de Janeiro, Brazil}
	\address{$^6$ Center for Computational Social Science, Kobe University, Kobe 657-8501, Japan.}
	
	\vspace{10pt}
	
	\begin{abstract} 
		Human mobility, contact patterns, and their interplay are key aspects of our social behavior that shape the spread of infectious diseases across different regions. In the light of new evidence and data sets about these two elements, epidemic models should be refined to incorporate both the heterogeneity of human contacts and the complexity of mobility patterns. Here, we propose a theoretical framework that allows accommodating these two aspects in the form of a set of Markovian equations. We validate these equations with extensive mechanistic simulations and derive analytically the epidemic threshold. The expression of this critical value allows us to evaluate its dependence on the specific demographic distribution, the structure of mobility flows, and the heterogeneity of contact patterns, {thus shedding light on the microscopic mechanisms responsible for the epidemic detriment driven by recurrent mobility patterns reported in the literature.}\\\\
		\textbf{Published version:} \\ \href{https://iopscience.iop.org/article/10.1088/1367-2630/ac0c99}{New J. Phys. \textbf{23}, 073019 (2021)} [DOI:10.1088/1367-2630/ac0c99]
	\end{abstract}
	
	
	\section{Introduction}
	
	The proliferation and accessibility of large data sets describing the essential aspects of human behavior is being crucial to reveal the influence that our social habits have on the development of epidemics as well as providing useful insights to design non-pharmaceutical containment strategies. Human mobility is one of the aspects of our social behavior determining the form and speed of the transmission of infectious diseases. In this sense, the recent availability of data about the mobility patterns of individuals at different levels~\cite{Guimera2005,gonzalez2008understanding,BARBOSA20181}, from global to urban, demands to revisit epidemic models, in particular those studying the geographical spread of pathogens leveraging the mobility of hosts \cite{BALL201563}.
	
	Data-driven models are developed to improve the spatio-temporal accuracy of predictions of real epidemic outbreaks by using a large amount of real data as inputs~\cite{eubank2004modelling,balcan2009seasonal,halloran2014ebola,bansal2016big,Zhang17,Kraemer20,Schlosser20}. However, agent-based and mechanistic models {based on large-scale stochastic Monte Carlo simulations} have as a counterpart the impossibility of performing analytical treatments that shed light on the role played by the different aspects of our sociability in the transmission of communicable diseases. To fill the gap between accurate epidemic forecasting systems and mathematical models, theoretical frameworks should be refined in order to be able to incorporate as much social data as possible. 
	
	The most usual way to incorporate mobility patterns into epidemic models is the use of metapopulations. In this case, individuals are considered to live in a set of subpopulations (or patches) whereas flows of individuals happen among these patches. Within this framework, the spread of diseases is characterized by local reactions inside each patch~\cite{Ball91,Sattenspiel95,Lloyd96,grenfell1997meta,Keeling02} that mimic the interactions between individuals giving rise to the transmission of the pathogen. This reaction process within each patch interplays with the global diffusion of agents that captures the mobility patterns at work. 
	
	The first metapopulation frameworks were built by considering assumptions that simplify their mathematical analysis while limiting their direct application in real situations. However, with the advent of the XXI century and the massive use of online platforms, real data capturing individual flows between different geographical areas were incorporated into metapopulation frameworks~\cite{colizza2007reaction,colizza2007invasion,colizza2008epidemic,balcan2009multiscale} in an attempt of increasing their accuracy while preserving the ability to perform analytical predictions.  Still, the first models in this line assumed simple mobility patterns such as random diffusion~\cite{mata2013meta,diogo2018mot} or continuous models of commuting flows~\cite{Simini2012,Simini2013,Masucci2013}, that allowed analytical studies about the influence of mobility on the epidemic threshold~\cite{colizza2007reaction}. 
	
	The next step in the search for more reliable and accurate metapopulation models was to get rid of the simplifying assumptions about human diffusion and find ways to take into account aspects such as the recurrent nature~\cite{balcan2011phase,belik2011natural,belik2011,balcan201287,gmezgardees2017} and high order memory of human displacements~\cite{Matamalas16}, the coexistence of different mobility modes~\cite{physrevx8031039}, or the correlation between the time-scales associated to human mobility and that of infection dynamics~\cite{soriano2020temporal}. These models, apart from yielding important insights about the role that human behavior has on the unfolding of epidemic states, have turned to be useful tools to reproduce the real prevalence distribution of endemic diseases \cite{soriano2020vector} and the advance of real epidemic outbreaks~\cite{Arenas2020,Costa2020br}, thus showing a versatile and hybrid facet as mathematical yet informative models.
	
	The former refinements have focused on the way real human mobility patterns are incorporated into metapopulation frameworks, but continue using simple mixing rules for the interaction of individuals within each patch. {These simplifying hypotheses include well mixing assumptions and explore scenarios where the number of contacts inside each patch is homogeneous and usually determined by some demographic aspects such as the density of the patch {or its age distribution.}} However, human contact patterns are  known to be highly heterogeneous and this attribute plays a central role in the transmission of some communicable diseases~\cite{SatorrasVespi2001}. In fact, the analysis of the propagation of recent coronavirus such as SARS-CoV-1~\cite{Shen04,Lloyd05,Stein11}, MERS-CoV~\cite{Wong15,Hui16} and SARS-CoV-2~\cite{Frieden20,MacKenzie20,Shim20,althouse2020,Suneabe2424,althouse2020stochasticity}, reveals that a small proportion of cases were responsible for a large fraction of the infections. This empirical  evidence supports the existence of super-spreading events \cite{Meyerowitz20}, an attribute of transmission chains that cannot be captured by models in which the contacts of individuals, and hence their infectiousness, are assumed to be homogeneous. 
	
	{There have been some attempts in the literature to account for the impact of individual diversity in metapopulation modeling~\cite{apolloni2014metapopulation,Arenas2020}. However, they usually rely on the stratification of the population into different age-groups~\cite{mistry2021inferring}, which are assumed to be homogeneous, and the introduction of mixing matrices governing the interactions among them. Therefore, a general formalism able to accommodate heterogeneous subpopulations with any arbitrary degree distribution is still missing in the literature.  In this paper, we aim at filling this gap and including the heterogeneity of social contact patterns in the body of a metapopulation model, in particular that presented in reference~\cite{gmezgardees2017} and used in subsequent works~\cite{physrevx8031039,soriano2020vector,Arenas2020}.} 
	
	{The most important result found in these works was the detrimental effect of human daily recurrent mobility for the emergence of epidemic outbreaks. Nonetheless, the mean-field assumption included within each subpopulation in these formalisms precludes getting any microscopic explanation about the mechanism triggering this phenomenon. The model presented here is therefore a step forward towards a metapopulation formalism that includes concomitantly the demographic distribution of real populations, the recurrent nature of human displacements, and the heterogeneity of social contacts and sheds light on the unexpected phenomena arising from their interplay. In fact,  the most important finding in this new framework is that the detrimental effect of human daily recurrent mobility is recovered despite  the  fact  that  the number  of  interactions does not depend on the number of agents that meet inside each patch. Thus, individual interactions appear here as an intensive parameter, rather than an extensive one as in  reference \cite{gmezgardees2017}, shedding light  on  the  microscopic roots of the epidemic detriment phenomenon.}
	
	\section{Metapopulation model}
	\label{sec:metapopmodel}
	
	\subsection{Coupling recurrent mobility and heterogeneous contacts}
	
	Let us start the construction of the metapopulation framework by describing the interaction rules that govern the mixing of individuals across and within patches. We consider a metapopulation network with $\Omega$ patches, each one of population $n_i$ ($i=1,\ldots,\Omega$), thus accumulating a total of $N =\sum_i n_i$ individuals. Each individual is associated with a single residence (one of the patches) and can travel to another patch according to some mobility rules. The flow of individuals from a patch $i$ to another $j$ is described by a directed and weighted network of patches, in which the weight $W_{ij}$ is the number of individuals from $i$ that commute to $j$ daily. The matrix $W_{ij}$ is also called origin-destination {(OD)} matrix and allows us to define the probability that, when an individual living in $i$ decides to move, she or he goes to patch $j$ as
	\begin{equation}
		\label{eq:1}
		R_{ij} = \ffrac{W_{ij}}{\sum_{l=1}^\Omega W_{il}},
	\end{equation}
	where $\sum_{l=1}^\Omega W_{il}=s_i$ is the total number of trips observed from patch $i$. 
	
	According to the framework presented in reference~\cite{gmezgardees2017}, mobility and interactions are iterated in consecutive rounds of a process that involves three stages: Mobility, Interaction, and Return (MIR). Namely, first the agents with residence in a patch $i$ decide to move with probability $p$ (or they stay in $i$ with probability $1-p$). If they move, their destination $j$ is chosen with probability $R_{ij}$, given by equation \eqref{eq:1}. Once all the agents in each patch have been assigned to their new locations (either their residence or a new destination chosen according to the matrix ${\bf R}$) the interaction on the assigned patch takes place with the rest of agents in the same subpopulation. Finally, once the interaction stage has finished, agents are placed in the original population, {\em i.e.}, they come back to their corresponding residence.
	
	\begin{figure}[t!]
		\centering
		\includegraphics[width=0.55\linewidth]{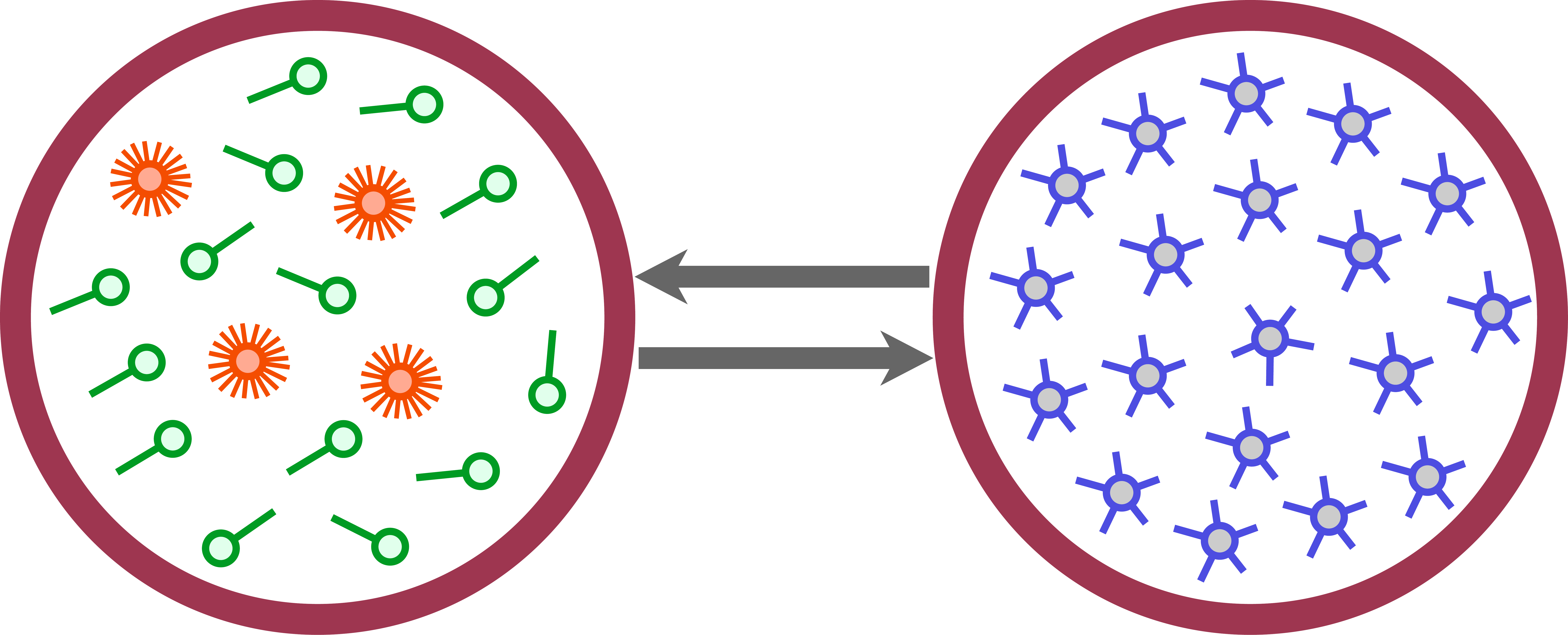}
		\caption[Example of a metapopulation with two patches]{Example of a metapopulation with two patches, both having the same average connectivity {$\av{k}=5$}. The first is a heterogeneous patch with resident individuals of connectivity $1$ or $20$, and the second is a homogeneous patch {in which all residents have} the same connectivity $5$.}
		\label{fig:rede2patches}
	\end{figure}
	
	Now we propose a modification to consider heterogeneous contacts inside each patch. In reference~\cite{gmezgardees2017}, all individuals inside a patch interact with all others with the same probability thus following a homogeneous mixing hypothesis. Here we propose a model in which each individual in a patch has a different social degree or \textit{connectivity} $k$ as shown in figure~\ref{fig:rede2patches}. In this way, each patch $i$ has $\nnik{i}{k}$ individuals with connectivity $k$, so that the population of patch $i$ can be written as:
	\begin{equation}
		n_i = \sum_k \nnik{i}{k} = \sum_k n_i P_i(k),
	\end{equation}
	where $P_i(k)$ is the probability that a randomly chosen individual living inside $i$ has a connectivity $k$:
	\begin{equation}
		P_i(k) = \ffrac{\nnik{i}{k}}{n_i}.
	\end{equation}
	
	In the following, we assume that individuals with social connectivity $k$ will preserve this value when traveling to another patch, {\em i.e.}, we assume that sociability is an intrinsic individual attribute that does not depend on their location. This later hypothesis captures the biological and behavioural aspect of hosts that can turn them into super-spreaders, {\em i.e.}, individuals that are highly efficient in transmitting the disease due to a high viral shedding \cite{Woolhouse338} or because they have a high contact rate due to a pronounced social behavior. However, other causes that are inherently related to the location, such as the existence of high-risk scenarios related to work or leisure, are not captured by the former assumption.
	
	Under the former hypothesis about the invariance of the connectivity $k$ under mobility and assuming that those individuals with connectivity $k$ move with probability $p_k$, we can calculate the effective population of a patch $i$, $\nnieff{i}$, after the movement stage has been performed, as the sum of the effective number of agents with connectivity $k$:
	\begin{equation}
		\label{eq:new_nieff}
		\nnieff{i} = \sum_k \nnikeff{i}{k}.
	\end{equation}
	In the latter equation, $\nnikeff{i}{k}$ is calculated considering the number of individuals with connectivity $k$ that travel from any patch $j$ to $i$:
	\begin{equation}
		\label{eq:new_nikeff}
		\nnikeff{i}{k} = \sum_j \nnik{j\to i}{k}, 
	\end{equation}
	where
	
	\begin{equation}
		\nnik{j\to i}{k}= \left[(1-p_k)\dij + p_k R_{ji}\right] n_j P_j(k).
	\end{equation}
	
	Another quantity that can be evaluated is the effective connectivity distribution of a patch, $\pokeff{i}(k)$, defined as the probability of finding an individual of connectivity $k$ in patch $i$ after the mobility stage. This probability is given by:
	
	\begin{equation}
		\label{eq:new_pokeff}
		\pokeff{i}(k) =  \ffrac{\nnikeff{i}{k}}{\nnieff{i}}\ .
	\end{equation}
	From the effective connectivity distribution of a patch $i$ we can measure the effective moments as:
	\begin{equation}
		\knieff{n}{i} = \sum_k k^n \pokeff{i}(k).
	\end{equation}
	
	\subsection{Disease spreading dynamics}
	
	The coupling of interaction and mobility patterns of agents produces, for a given set of mobility probabilities $\{p_k\}$, a variation of the main structural attributes of the patches, as shown by the expressions of the effective population, equations \eqref{eq:new_nieff}-\eqref{eq:new_nikeff}, and  the effective connectivity distribution, equation \eqref{eq:new_pokeff}. These variations occur once the mobility step is performed and become crucial when the spreading process (the interaction step of the MIR model) enters into play.
	
	Here the interaction stage is incorporated as a Susceptible-Infected-Susceptible (SIS) spreading dynamics. To this aim, we denote the number of infected individuals residing in $i$ that have connectivity $k$ as $\Iik{i}{k}$, implying that the total number of infected residents in $i$ is $I_i = \sum_k \Iik{i}{k}$. Thus, the probability that an agent with residence in patch $i$ and connectivity $k$ is infected is given by:
	\begin{equation}
		\rinfk{i}{k} = \ffrac{\Iik{i}{k}}{\nnik{i}{k}}\;.
		\label{eq:rinfik}
	\end{equation}
	The probabilities $\{\rinfk{i}{k}\}$ (with $i=1,\ldots,\Omega$ and $k=1,\ldots,\kmax$) constitute our dynamical variables. From these variables we can compute the fraction of infected individuals with residence in patch $i$: 
	\begin{equation}
		\rho_i = \sum_k  \rinfk{i}{k}P_i(k)\;,
	\end{equation} 
	or the fraction of infected individuals in the whole metapopulation:
	\begin{equation}
		\rho = \frac{1}{N} \sum_i {n_i} \rho_{i}\;.
	\end{equation}
	
	To derive the corresponding Markovian evolution equations of the probabilities $\{\rinfk{i}{k}\}$  corresponding to the SIS dynamics we make use of the so-called heterogeneous mean-field theory (HMF) in the annealed regime \cite{RMP15}. Thus, after the movement stage, each susceptible agent with connectivity $k$ that is placed in patch $j$ connects randomly with $k$ individuals in the same patch and, for each infected contact, the susceptible agent will become infected and infectious with probability $\blb$. In addition, those infected agents at time $t$ will recover and become susceptible again with probability $\bmu$. Following these simple rules, the equations for the time evolution of the probabilities $\{\rinfk{i}{k}\}$ read:
	\begin{equation}
		\label{eq:new_eq}
		\rinfk{i}{k}(t+1) = (1-\bmu) \rinfk{i}{k}(t) + \left[    1 - \rinfk{i}{k}     \right]{\pik{i}{k}(t)},
	\end{equation}
	where $\pik{i}{k}(t)$ is the probability that a healthy individual with connectivity $k$ {and} residence in patch $i$ becomes infected at time $t$:
	\begin{equation}
		\label{eq:new_piik}
		\pik{i}{k}(t) = (1-p_k) \Qik{i}{k}(t) + p_k  \sum_{j=1}^\Omega R_{ij} \Qik{j}{k}(t)\;,
	\end{equation}
	where $\Qik{i}{k}(t)$ is the probability that an individual of connectivity $k$ placed in patch $i$ becomes infected at time $t$ and reads: 
	\begin{equation}
		{
			\label{eq:new_Qik}
			\Qik{i}{k}(t) = 1 - \left(      1- \blb\sum\limits_{k'}{\pokeff{i}(k'|k)} {\rinfkeff{i}{k'}(t)} \right)^k\;.
		}
	\end{equation}
	In the former expression, $\pokeff{i}(k'|k)$ is the probability that an agent with connectivity $k$ placed in patch $i$ is connected with another agent with $k'$ {placed} in the same patch. In addition, $\rinfkeff{i}{k}$ is the effective fraction of infected individuals with connectivity $k$ placed in patch $i$:
	\begin{equation}
		\label{eq:rinfeff}
		\rinfkeff{i}{k} = \frac{\Iikeff{i}{k}}{\nnikeff{i}{k}} =  \frac{1}{\nnikeff{i}{k}}\sum_j \Iik{j\to i}{k}=
		\frac{
			1
		}
		{
			\nnikeff{i}{k}
		}
		\sum_{j} \nnik{j\to i}{k} \rinfk{j}{k}(t)\;,
	\end{equation}
	where the denominator is given by \eqref{eq:new_nikeff} and the numerator is the number of infected individuals that  are in patch $i$. 
	
	In the following we will consider that the contact networks created at each interaction step are completely uncorrelated. This way, the probability $\pokeff{i}(k'|k)$ can be written in terms of the effective connectivity distribution of patch $i$ as:
	\begin{equation}
		\label{eq:new_pklk_eff}
		\pokeff{i}(k'|k)  = \ffrac{k' \pokeff{i}{(k')}}{\knieff{}{i}}=\ffrac{k' \nnikeff{i}{k'}}{\sum_{k^{''}} k^{''} \nnikeff{i}{k^{''}} }\;, 
	\end{equation}
	which is the probability of selecting an edge from an individual with connectivity $k'$ placed in patch $i${, independent of $k$}.
	
	\section{Metapopulations with heterogeneous subpopulations}
	\label{sec:metapopk_results}
	
	{The derived Markovian equations  are general for a set of $\Omega$ patches, their population $n_i$, degree distribution $P_i(k)$, and OD matrix elements $W_{ij}$, ($i,j=1,\ldots,\Omega$). We} now study the impact of heterogeneous distributions of individual contacts by using synthetic metapopulations to validate the{se} equations by comparing the results obtained by the iteration of equations \eqref{eq:new_eq}-\eqref{eq:new_Qik} with the results of mechanistic Monte Carlo (MC) simulations in which we keep track of the dynamics of each  agent.
	
	\begin{figure}[t!]
		\centering
		\includegraphics[width=0.55\linewidth]{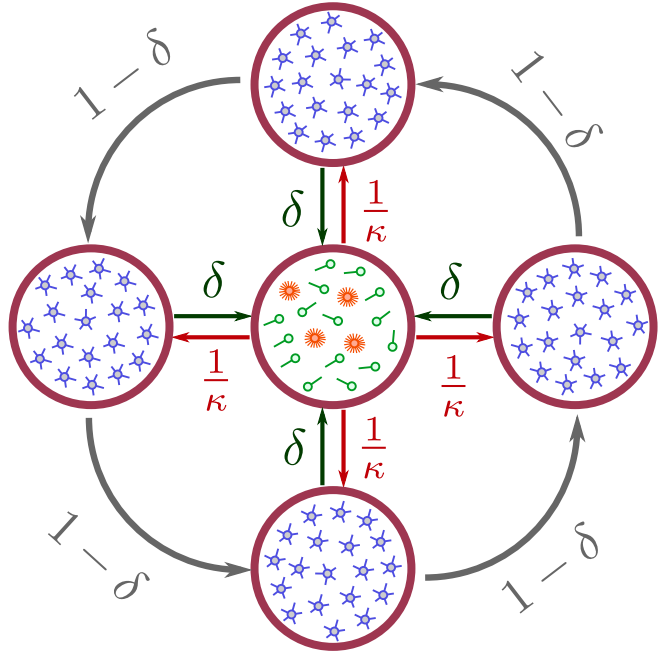}
		\caption{Example of a star-like metapopulation network with $\kappa + 1$  patches. In this example, the leaves and the hub have the same number of individuals, $n_l = \alpha n_h$, with $\alpha = 1$, while the hub is a heterogeneous patch with resident individuals of connectivity 1 with probability $\eta$, or $\kmax = 20$ with complementary probability $1 -\eta$, and each leaf is a homogeneous patch with residents of same connectivity $\av{k}_l =  \beta\av{k}_h$, with $\beta = 1$ and $\av{k}_h = 5$. The flow from hub to a leaf happens with probability $R_{hl} = \kappa^{-1}$, from leaves to hub with $R_{lh} = \delta$, and between adjacent leaves with $R_{l,l+1} = 1-\delta$, in counterclockwise direction.}
		\label{fig:rederoda}
	\end{figure}
	
	\subsection{Synthetic metapopulation}
	\label{sub:star}
	
	Although the formalism presented can accommodate any arbitrary mobility network and set of connectivity distributions, we restrict our analysis, as in reference~\cite{gmezgardees2017}, to synthetic star-like metapopulation networks. Our choice is rooted in their versatility for, despite being simplistic  structures,  star-like metapopulations exhibit a wide variety of regimes caused by the non-uniform distribution of the population across patches and the asymmetry in th mobility patterns connecting them. This kind of synthetic metapopulation, shown in figure~\ref{fig:rederoda}, is composed by a central patch (the hub) connected to $\kappa$ patches (the leaves). The hub $h$ has a population of $n_h$ individuals, while each leaf $l$ has a fraction $\alpha \in [0,1]$ of the hub population, $n_l = \alpha n_h$. The mobility towards leaves of individuals with residence in the hub is uniform, given by:  
	\begin{equation}
		R_{hl} = \ffrac{1}{\kappa},
		\label{eq:starRhl}
	\end{equation}
	while the mobility of those residents in the leaves is controlled by a parameter $\delta$. This way, a resident in a leave $l$ that decides to move will go to the hub with probability $\delta$,
	\begin{equation}
		R_{lh} = \delta,
		\label{eq:starRlh}
	\end{equation}
	or move to the next (counterclockwise direction) leave with probability
	\begin{equation}
		R_{l,l+1} = 1-\delta.
		\label{eq:starRllp1}
	\end{equation}
	Note that the choice of the direction of movements among leaves is not relevant as long as it is uniform across all the leaves, for they are statistically equivalent. Up to this point, the design of the metapopulation is identical to that presented in reference~\cite{gmezgardees2017}, being characterized by two parameters $\alpha$ and $\delta$. However, the synthetic metapopulations used here {get rid of} the assumption of homogeneous (all-to-all) contact patterns in the patches. To this aim, and keeping the symmetry of the original star-like metapopulations, we consider that the residents of the central patch (the hub)  have a contact distribution $\pok{h}(k)$ that is different from that of the residents in the leaves, $\pok{l}(k)$. A particular case of this setting used along the manuscript is to consider that the connectivity distribution of the individuals belonging to the hub is bimodal: 
	\begin{equation}
		\pok{h}(k) = \eta \delta_{k1} + (1-\eta)\delta_{k\kmax},
	\end{equation}
	{\em i.e.}, agents in the hub have connectivity $1$ with probability $\eta$ and connectivity $\kmax$ with probability $(1-\eta)$. This way,
	the $n$-th moment of the hub's connectivity distribution is:
	\begin{equation}
		\av{k^n}_h=\sum_k k^n\pok{h}(k) = \eta + (1-\eta)\kmax^n\;.
	\end{equation}
	In their turn, those individuals belonging to leaves have the same number of contacts ($\av{k}_l$):
	\begin{equation}
		\pok{l}(k) = \delta_{k\av{k}_l}\;.
	\end{equation}
	
	Note that the values of $\eta$ and $\kmax$ are correlated if we impose the additional constraint that the hub has an average connectivity $\av{k}_h$ fixed. In this case, given a value $\kmax$, the value of $\eta$ that allows it is given by:
	\begin{equation}
		\eta = \frac{\kmax-\av{k}_h}{\kmax-1}.
		\label{eq:eta}
	\end{equation}
	
	In this simple configuration, the heterogeneous nature of the contacts is two-fold. From a microscopic point of view, the bimodal distribution existing inside the central node induces local heterogeneities in the contacts made by residents there, which are controlled by parameters $\eta$ and $\kmax$. In its turn, another global connectivity heterogeneity emerges driven by the asymmetry existing between the connectivity of residents of the hub and the leaves. In particular, we will assume throughout the manuscript that $\av{k}_l = \beta \av{k}_h$, with $\beta \in[0,1]$. According to this formulation, the star-like metapopulation shown in figure~\ref{fig:rederoda} has $\av{k}_h=5$, $\kmax = 20$, and $\alpha = \beta = 1$.
	
	\subsection{Monte Carlo simulations}
	
	To check the validity of the Markovian equations, we define a MC algorithm for the stochastic simulation of the SIS model on top of a metapopulation with heterogeneous contact patterns. As in the case of Markovian equations, equations \eqref{eq:new_eq}-\eqref{eq:new_Qik},  the  proposed process  is also a discrete-time dynamics. At each time step $t$, each individual is tested to move with probability $p_k$ (being $k$ the number of contacts assigned to this individual). If accepted, it moves to a patch $j$ with probability $R_{ij}$. Then, each susceptible individual with connectivity $k$ chooses randomly $k$ individuals in the patch they currently occupy and are infected with probability $\blb$ if the contacted individual is infectious. Once all the potential infections events have been simulated, healing happens with probability $\bmu$ for each infected individual at time {$t-1$. In this sense, we perform a synchronous update of the state of the entire metapopulation.} 
	
	{First, a fraction $\rho_\mathrm{ini}$ of the population is randomly infected as the initial condition and the} simulation procedure in a give time step $t$ can be summarized as follows:
	\begin{enumerate}
		\item\label{step:init}  For each patch $i$, each individual with connectivity $k$  resident in $i$ is tested to move with probability $p_k$. If she or he moves, a patch $j$ is chosen  proportionally to $R_{ij}$. 
		\item\label{step:mcnew} Each susceptible individual with connectivity $k$ selects {$k$ contacts} at random in patch $i$. For each attempt, it can be infected with probability: 
		\begin{equation}
			\blb \ffrac{\sum_k k \Iikeff{i}{k}}{\sum_k k \nnikeff{i}{k}}\;,\end{equation} 
		or remains susceptible with complementary probability. These attempts stop when the individual becomes infected {and reproduce the annealed regime proposed in section \ref{sec:metapopmodel}, since all edges are available for each individual in the same time step}.
		\item Each individual with infected state at time step $t-1$ heals in time step $t$ with probability $\bmu$.
		\item Finally, all individuals return to their residences and time step $t+1$ starts in (\ref{step:init}).
	\end{enumerate}
	
	To avoid the absorbing state, we infect a small fraction $\rho_\mathrm{pump} = 2\times 10^{-4}$ of individuals at random when this state is reached~\cite{SanderQS,Cota2017}. This keeps the dynamics always active and the equilibrium state is defined after comparing averages over sequential time windows of size $T = 100$, and accepting if the absolute difference is smaller than {$\rho_\textrm{cvg} = 10^{-6}$}.
	
	\subsection{Comparison between MC and Markovian equations}
	
	The comparisons between MC and Markovian equations are performed in star-like metapopulations with $\kappa = 10$ and $\alpha=1$, {\em i.e.}, in which all patches (hubs and leaves) contain the same number of individuals ($n_l = n_h=10^4$ individuals per patch),  to focus on the effect of contact heterogeneity. Furthermore, for the same reason, we focus on the case that mobility is independent of the connectivity of individuals, $p_k=p~\forall~k$.
	
	\begin{figure}[t!]
		\centering
		\includegraphics[width=\linewidth]{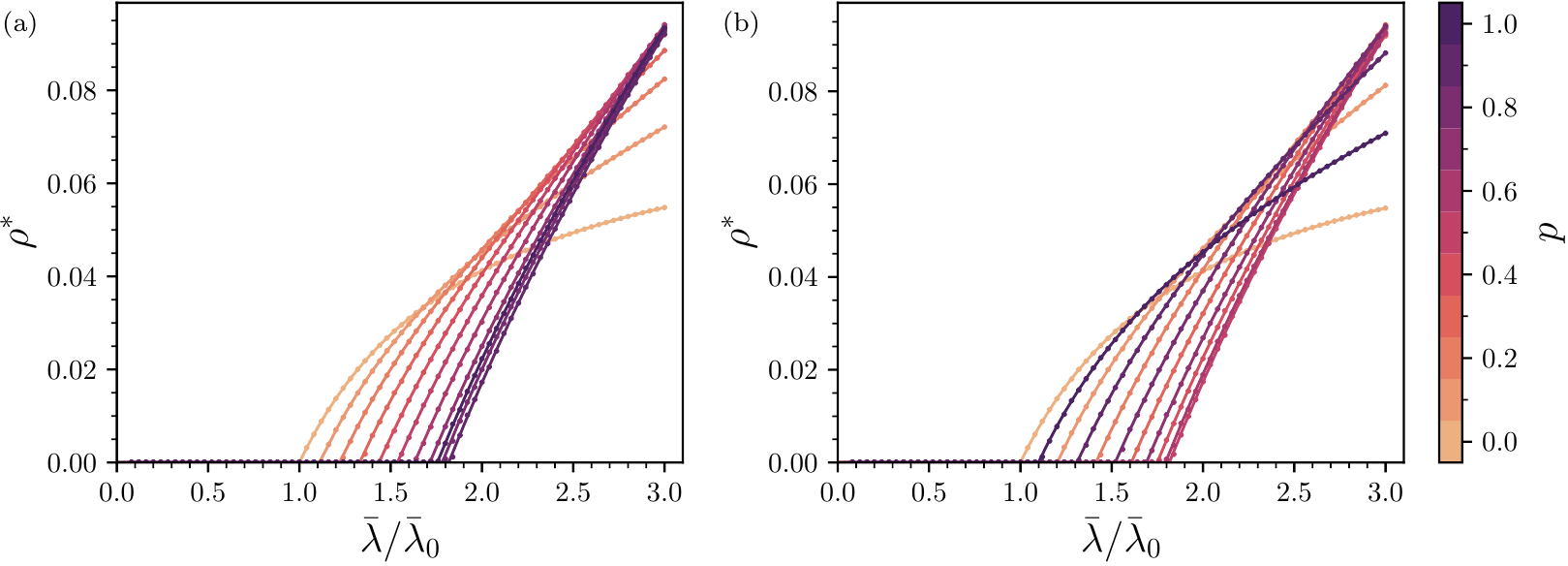}
		\caption{Equilibrium regimes of the Markovian equations (lines) and MC simulations (symbols) for a star-like metapopulation with $n_h = n_l= 10^4$ and $\kappa = 10$. The hub contains individuals with connectivity $\av{k}_h = 100$ ($\eta = 0$ and $\kmax = 100$), and the leaves $\av{k}_l = 10$ ($\beta = 0.1$). The mobility patterns are given by (a) $\delta = 0.1$ and (b) $0.9$.  A fraction $\rho_\mathrm{pump}= 2\times 10^{-4}$ and ten stochastic samples were used for the MC simulations.}
		\label{fig:MCnew-star1}
	\end{figure}
	
	First {we neglect local heterogeneities} and consider that contact heterogeneity only happens between patches. {In mathematical terms, this assumption} implies that the population of the hub has an homogeneous contact distribution ($\eta=0$) although its mean connectivity $\av{k}_h=\kmax$ is {different from that of the leaves $\av{k}_l= \beta \av{k}_h$, with $\beta \neq 1$.} In particular, in figure~\ref{fig:MCnew-star1} we plot the mean epidemic prevalence $\rho^*$ in the equilibrium state as a function of the infection probability $\blb$ scaled by the epidemic threshold in the case of null mobility {$\blb_0\equiv\blbc(p=0)$}. {To derive the latter quantity, we realize that the absence of flows among the patches precludes the interaction among the residents in different areas, so the epidemic threshold corresponds to the well-known expression provided by HMF equations \cite{RMP15} for the most vulnerable patch. Therefore,
		\begin{equation}
			\blb_0 = \bmu \min\left\lbrace\frac{\av{k}_h}{\av{k^2}_h},\frac{\av{k}_l}{\av{k^2}_l}\right\rbrace\ .
	\end{equation}}
	We consider that $\av{k}_h=100$ while leaves have $\av{k}_l=10$ ($\beta = 0.1$) and explore two different mobility patterns. In particular, in (a) we set $\delta=0.1$ so that most of the residents of leaves move circularly, {\em i.e.}, passing from one leave to another and avoiding the hub. In this case, the so-called epidemic detriment by mobility shows up so that the epidemic state is delayed as the mobility $p$ increases, with the exception of very large values of $p$. However, note that, at variance with reference~\cite{gmezgardees2017}, here {both the hubs and the leaves are equally populated}; we will explore the roots of this detriment below. Second, in panel (b), we set $\delta=0.9$ so that the situation is the opposite and the residents of leaves tend to visit the hub. In this case, the epidemic detriment is also evident although this behavior is restricted to values $p<0.5$, while for $p>0.5$ the increase of mobility produces a progressive decrease of the epidemic threshold. In both cases, the agreement with MC simulations is almost perfect.
	
\begin{figure}[t!]
		\centering
		\includegraphics[width=\linewidth]{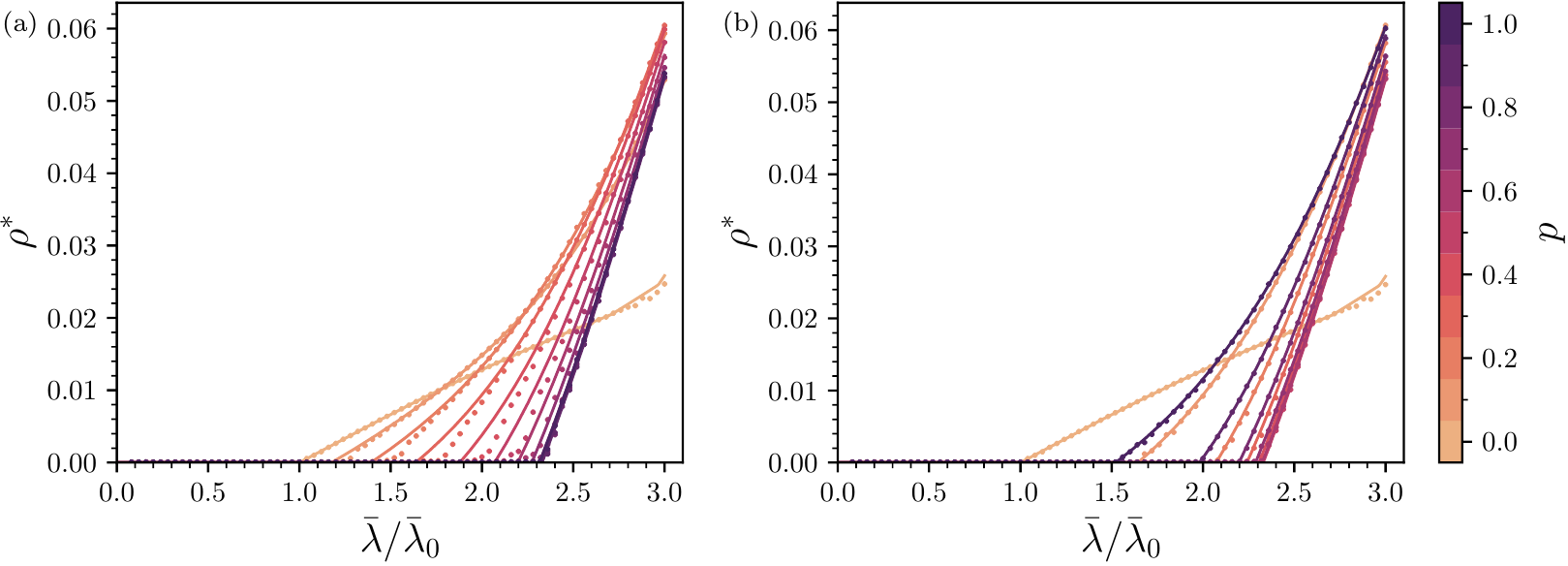}
		\caption{Equilibrium regimes of the Markovian equations (lines) and MC simulations (points) for $n_h=n_l = 10^4$ and $\kappa = 10$.  The patches contain individuals with power-law connectivity distributions $P_i(k)\sim k^{-\bar\gamma_i}$, $k\in[3,100]$, with $\bar\gamma_h = 2.3$ for the hubs and $\bar\gamma_l = 3.5$ for the leaves. The mobility pattern is given by (a) $\delta = 0.1$ and (b) $0.9$.  A fraction $\rho_\mathrm{pump}= 2\times 10^{-4}$ and ten stochastic samples were used for the MC simulations.}
		\label{fig:MCnew-star2}
	\end{figure}

	Next we analyze a star-like metapopulation that generalizes the contact heterogeneity of the first one. In this case the hub is very heterogeneous, containing a power-law distribution, $P_h(k)\sim k^{-\bar\gamma_h}$ with $\bar\gamma_h = 2.3$, while leaves have also a power-law distribution $P_l(k)\sim k^{-\bar\gamma_l}$ with  $\bar\gamma_l=3.5$, both with $k \in [3,100]$, the hub being the most heterogeneous one. The cases explored in figure~\ref{fig:MCnew-star2} are again (a) $\delta = 0.1$ and (b) $\delta=0.9$, showing similar qualitative behaviors with the mobility, namely the emergence of epidemic detriment, to those found in figure~\ref{fig:MCnew-star1}.
	Quantitatively, it is worth stressing that the existence of strong local heterogeneities within both hub and leaves  in absence of mobility will lead to an activation described by the HMF theory, in which the epidemic prevalence approaches zero close to the epidemic threshold as $\rho\sim (\blb-\blbc)^{\bar\beta}$ where $\bar\beta>1$ if the degree exponent is smaller than 4~\cite{SatorrasVespi2001pre}, and valid for large population sizes (thermodynamic limit). The convexity of the prevalence curve approaching the transition in the finite-size population of the investigated patches is reminiscent of this behavior. Again, the agreement with MC is good, except around the epidemic threshold due to difficulties in avoiding the absorbing state.
	
	\section{Epidemic threshold}
	
	Figures \ref{fig:MCnew-star1} and \ref{fig:MCnew-star2} reveal that the epidemic detriment emerges even when dealing with {uniformly distributed populations, contrarily} with reference~\cite{gmezgardees2017}, in which increasing mobility in homogeneous populations favors epidemic spreading by {reducing} the epidemic threshold, $\blbc$, here defined as as the minimum infectivity per contact, $\blb$, such that an epidemic state can be stable. Therefore, the emergence of epidemic detriment here should be rooted in the interplay among contact heterogeneities and human mobility. In this section, we aim at deriving an analytical expression of the epidemic threshold, $\blbc$ {for general configurations}, to shed light on the mechanisms giving rise to the behavior shown above. 
	
	Let us assume that the dynamics has reached its steady state, so that $\rinfk{i}{k}(t+1)= \rinfk{i}{k}(t)= \rinfkst{i}{k}$. Under this assumption, equation~\eqref{eq:new_eq} reads:
	\begin{equation}
		\bmu\rinfkst{i}{k} = \left[    1 - \rinfkst{i}{k}\right]\pikst{i}{k}  
		\label{eq:threst}
	\end{equation}
	with 
	\begin{equation}
		\pikst{i}{k}  = (1-p_k) \Qikst{i}{k} + p_k  \sum_{j=1}^\Omega R_{ij} \Qikst{j}{k}\ .
		\label{eq:pist}
	\end{equation}
	Furthermore, for $\blb$ values close to the epidemic threshold, the fraction of infected individuals is negligible, which means that $\rinfkst{i}{k} =\epik{i}{k}\ll 1~\forall~(i,k)$. This fact allows us to linearize the equations characterizing the steady state of the dynamics by neglecting all the terms $\mathcal{O}(\bar\epsilon^2)$. In particular, the probability that an individual with connectivity $k$ and placed in $i$ contracts the disease, $\Qikst{i}{k}$, can be approximated by
	\begin{equation}
		{
			\Qikst{i}{k} = 1 -  \left(   1 - \blb\sum\limits_{k'}    \pokeff{i}(k'|k) \rinfkeffst{i}{k'}    \right)^k  \simeq \blb k  \sum_{k'} \pokeff{i}(k'|k) \rinfkeffst{i}{k'}\ ,
		}
	\end{equation}
	where we have used ${\mathcal O}(\tilde{\rho}) = {\mathcal O}(\bar\epsilon)$ as shown by equation~\eqref{eq:rinfeff}. In particular, plugging \eqref{eq:rinfeff}-\eqref{eq:new_pklk_eff} into the last expression leads to:
	\begin{eqnarray}
		\label{eq:new_Qikst}
		\Qikst{i}{k}  
		&= \ffrac{\blb k}{Q_i} \sum_{k'} k' \sum_{j} \left[  \left(1 - p_{k'}\right)\dij  + p_{k'} R_{ji} \right] \nni{j} \pok{j}({k'}) \epik{j}{k'},
	\end{eqnarray}
	where
	\begin{equation}
		Q_i \equiv \sum_{k} k
		\sum_j \left[  (1-p_k)\dij + p_k R_{ji} \right] n_j \pok{j}(k)
		\label{eq:qthresholdst}	
	\end{equation}
	is the effective number of edges in patch $i$. Note that $\sum_i Q_i 
	= \sum_{k}\sum_{j} k \pok{j}(k) n_j$ is the total number of edges in the system,  a conserved quantity. After introducing \eqref{eq:new_Qikst} and some algebra, equation~\eqref{eq:pist} transforms into: 
	\begin{equation}
		\label{eq:pikst}
		\pikst{i}{k} = \blb \sum_j \sum_{k'} \bMijkkl
		\epik{j}{k'},
	\end{equation}
	where
	\begin{eqnarray}
		\bMijkkl = k k' \pok{j}(k') \left[ 
		{(1-p_k)(1-p_{k'}) \ffrac{\dij}{Q_i}}
		+ 
		{(1-p_k)p_{k'} \ffrac{R_{ji}}{Q_i}}
		+ 
		\nonumber \right.  \\ \left. \qquad\qquad\qquad\qquad\qquad\qquad
		{p_k(1-p_{k'}) \ffrac{R_{ij}}{Q_j}}
		+ 
		{p_kp_{k'} \sum_l \ffrac{R_{il}R_{jl}}{Q_l}}
		\right] n_j.
		\fl\label{eq:new_matrizMAntes}
	\end{eqnarray}
	
	Finally, if we introduce these values into equation~\eqref{eq:threst} and retain only linear terms in {$\bar\epsilon$}, we arrive to the following expression
	\begin{equation}
		\label{eq:new_eigeneqAntes}
		{
			\bmu \epik{i}{k} = \blb \sum_j \sum_{k'} \bMijkkl
			\epik{j}{k'}
		},
	\end{equation}
	that defines an eigenvalue problem. According to its definition, the epidemic threshold is thus given by:  
	\begin{equation}
		\blbc = \ffrac{\bmu}{\Lambda_\mathrm{max}({\bf {\bar M}})}.
		\label{eq:thresholdcomplete}
	\end{equation}
	The elements of matrix  ${\bf {\bar M}}$ given by \eqref{eq:new_matrizMAntes} represent four types of interactions in the metapopulation. Namely, the element $\bMijkkl$ represents the probability that a resident of patch $i$ with connectivity $k$ is in contact with another individual of patch $j$ and connectivity $k'$.  The first term accounts for interactions of residents of the patch, that do not move. In  second term, an individual of $i$ stays and interacts with a traveler from patch $j$ in patch $i$, that arrived with probability $p_{k'}R_{ji}$. A similar event happens in the third term, in which an individual of $i$ travels to patch $j$ and interact there with a resident of $j$ with probability $p_{k}R_{ij}$. Finally, in the forth term, both individuals of patches $i$ and $j$ travel to a patch $l$, arriving there with probability $p_kp_{k'}R_{il}R_{jl}$. {In computational terms, each row or column identifies individuals from one degree class living inside a patch. Therefore, the dimension of the matrix corresponds with the sum of the different degree classes observed within each patch}.
	
	\subsection{Homogeneous mobility across degree classes}
	
	{Equation~\eqref{eq:thresholdcomplete}} computes the exact expression of the epidemic threshold in presence of heterogeneous contact patterns. However, its computation involves solving the spectrum of a matrix whose dimension is determined by the number of connectivity classes and patches in the metapopulation. In particular, in presence of highly heterogeneous populations with fine spatial resolution, this problem can be computationally very hard due to a large number of elements of the critical matrix. For this reason, in what follows, we assume that mobility is independent of the connectivity so that $p_k = p$ which will considerably reduce the complexity of the problem as proved below.
	
	Before going ahead, it is convenient to make the transformation $\epik{i}{k}  \mapsto  k \bepik{ik}$ in equation~\eqref{eq:new_eigeneqAntes}. {Note that this represents a similarity transformation which does not alter the spectrum of the matrix. After doing such transformation, equation~\eqref{eq:new_eigeneqAntes} turns into}
	\begin{equation}
		\label{eq:new_eigeneq}
		{
			\bmu \bepik{ik} = \blb \sum_j \sum_{k'} \Mijkkl
			\bepik{jk'}
		},
	\end{equation}
	where the elements of the new matrix {$\bf M$} read as
	\begin{eqnarray}
		\Mijkkl = k'^2 \pok{j}(k') \left[ 
		{(1-p_k)(1-p_{k'}) \ffrac{\dij}{Q_i}}+ {(1-p_k)p_{k'} \ffrac{R_{ji}}{Q_i}}+ 
		\nonumber \right. \\
		\left. \qquad\qquad\qquad\qquad\qquad\qquad
		{p_k(1-p_{k'}) \ffrac{R_{ij}}{Q_j}}+ 
		{p_kp_{k'} \sum_l \ffrac{R_{il}R_{jl}}{Q_l}}
		\right] n_j.
		\label{eq:new_matrizM}
	\end{eqnarray}
	
	\noindent If $p_k = p$, equation~\eqref{eq:new_eigeneq} becomes independent of $k$, which allows a dimensionality reduction of the matrix. In particular, equation  \eqref{eq:new_eigeneq} reads:
	\begin{equation}
		\label{eq:new_eigeneqind}
		{
			\bmu \bepik{i} = \blb \sum_j M_{ij}\ 
			\bepik{j}
		},
	\end{equation}
	and the elements of the reduced matrix ${\bf M}$ are given by:
	\begin{equation}
		\Mij = \av{k^2}_j \left[ 
		{(1-p)^2 \ffrac{\dij}{Q_i}}+ p(1-p) \left( \ffrac{R_{ji}}{Q_i}  +  \ffrac{R_{ij}}{Q_j} \right)+ 
		{p^2 \sum_l \ffrac{R_{il}R_{jl}}{Q_l}}
		\right] n_j,
		\label{eq:new_matrizMind}
	\end{equation}
	where the effective number of edges $Q_i$ is now expressed as
	\begin{equation}
		Q_i = \sum_j  \av{k}_j \left[ (1-p)\dij + p R_{ji} \right]n_j\;.
	\end{equation}
	Once matrix ${\bf M}$ is constructed the epidemic threshold is computed as
	\begin{equation}
		\blbc = \ffrac{\bmu}{\Lambda_\mathrm{max}({\bf M})}.
		\label{eq:newlambdac}
	\end{equation}
	
\begin{figure}[t!]
		\centering
		\includegraphics[width=1.0\linewidth]{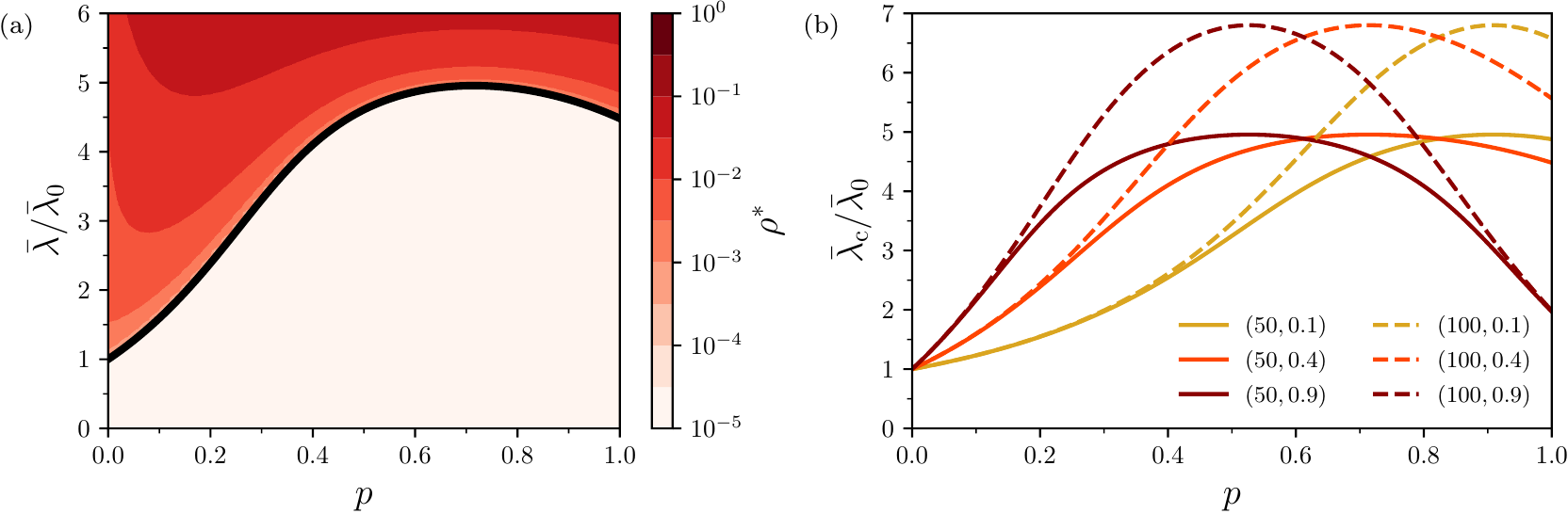}
		\caption{Dependence of the epidemic threshold on the mobility parameter $p$. All  patches have the same population ($\alpha = 1$), where the hub has agents with a bimodal connectivity distribution with $k = 1$ or $\kmax$, fixing $\av{k}_h = 5$, and the leaves have agents with connectivity $\av{k}_l = 5$ ($\beta = 1$). (a) Comparison of the theoretical epidemic threshold obtained using equation~\eqref{eq:newlambdac} (solid line), scaled by its value for $p = 0$, and the steady values of the prevalence $\rho$ obtained from equations \eqref{eq:new_eq}-\eqref{eq:new_Qik} {for $(\kmax,\delta) = (50,0.4)$}. (b) Relative epidemic threshold for different configurations $(\kmax,\delta)$, shown in the legends, with solid and dashed lines for $\kmax = 50$ and $100$, respectively.}
		\label{fig:threshold-curves}
	\end{figure}

	To test the accuracy of the former expression for the epidemic threshold, we compare its value computed according to equation~\eqref{eq:newlambdac} with the heat map of the steady state of the dynamics obtained from the iteration of equations \eqref{eq:new_eq}-\eqref{eq:new_Qik}. Figure~\ref{fig:threshold-curves}(a) reveals that the theoretical prediction of the epidemic threshold by equation~\eqref{eq:newlambdac} is very accurate and captures the dependence of the epidemic threshold on the mobility $p$. This threshold increases while promoting mobility until it reaches a maximum at $p=p^*$ since the infection is gradually reduced in the hub as $p$ increases, and the activation is then triggered in the leaves since hub's residents spend longer times there.
	
	For the sake of completeness, in \ref{app:new-static}, we analyze the case $p=0$ for equation~\eqref{eq:newlambdac} retrieving, as expected, the expression for the epidemic threshold provided by HMF equations on contact networks. Moreover, to quantify the effects of promoting mobility among disconnected patches, we perform a perturbative approach to the latter threshold which holds for small $p$ values in \ref{app:new-firstSecond}. Interestingly, at variance with the perturbative analysis carried out for (non-structured) well mixed metapopulations in reference~$\cite{gmezgardees2017}$, here the linear correction of the epidemic threshold strongly depends on the topological properties of the metapopulation. 
	
	\subsection{Disentangling the roots of the epidemic detriment}
	
	In what follows, to shed light on the nature of the epidemic detriment, we aim at quantifying the impact of the different components of the formalism, namely the underlying metapopulation structure and the contact heterogeneities existing among its population, on the relative magnitude $\blbc(p^*)/\blb_{0}$. To simplify this analysis, we will focus on the case of mobility independent of $k$, $p_k=p$, and consider the configuration defined in section~\ref{sub:star}, in which the hub has individuals with connectivity either $1$ or $\kmax$, with fixed average connectivity $\av{k}_h$, and the ones of the leaves have the same connectivity $\av{k}_l = \beta \av{k}_h$. For the sake of clarity, let us also express $\av{k^2}_l = \gamma \av{k^2}_h$. Note that in this configuration the values of $\eta$ and $\kmax$ are correlated by equation~\eqref{eq:eta}, while $\gamma$ is also correlated with $\beta$ and $\kmax$ via
	\begin{equation}
		\gamma = \frac{\beta^2 \av{k}_h^2}{\av{k}_h \left( \kmax + 1 \right) - \kmax }.
		\label{eq:gamma}
	\end{equation}
	
	\begin{figure}[t!]
		\centering
		\includegraphics[width=\linewidth]{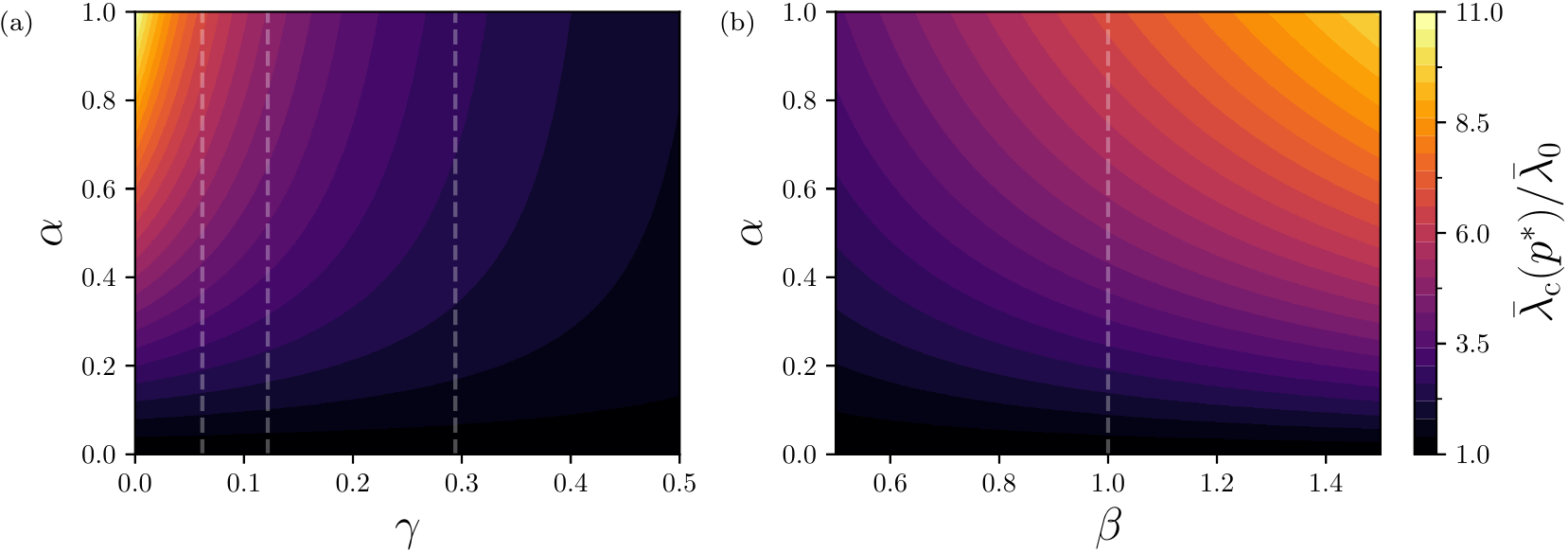}
		\caption{Heat maps of the relative magnitude of the peak of the epidemic threshold $\blbc(p^*)/\blb_0$ as a function of $\alpha$, $\beta$, and $\gamma$, with $\av{k}_h = 5$. In (a), all patches have the same average connectivity, with $\beta = 1$ while the local heterogeneity of the hub is modulated by $\gamma$. Dashed lines correspond to the values of $\gamma$ for $\kmax = 100$, $50$, and $20$, from left to right. The plot in (b) considers a fixed value of $\gamma \approx 0.0617$, corresponding to $\kmax = 100$ when $\beta = 1$, tuning the  connectivity of the leaves with  $\beta$.
			The population asymmetry is modulated by $\alpha$ for all cases.
		}
		\label{fig:phasediagram}
	\end{figure}
	
	First, we fix $\alpha = \beta = 1$, so that $n_l = n_h$ and $\av{k}_h = \av{k}_l$,  to study the effects of varying either the local heterogeneity existing in the hub by tuning $\kmax$ or the flows from leaves to the hub with $\delta$ in figure~\ref{fig:threshold-curves}(b). Fixing $\kmax = 50$ and changing $\delta$, it becomes clear that the increase of $\delta$ leads to a decrease of $p^*$ as a consequence of the higher mixing among individuals from the central node and the leaves, but does not change the relative magnitude $\blbc(p^*)/\blb_{0}$. 
	
	The former beneficial effect is rooted in the homogenization of the connectivity distribution driven by the mixing among individuals from the hub and the leaves. Interestingly, the position of the peak $p^*$ remains unaltered when keeping $\delta$ constant. Moreover, for small values of $p$, the behavior does not depend on the local heterogeneities of the patches, as shown by a perturbative analysis in \ref{app:new-firstSecond}.	Quantitatively,  it becomes clear that increasing the degree heterogeneity in the central node boosts the beneficial effect of the mobility, since the homogenization effect gains more relevance due to the higher vulnerability of the central node. {Mathematically, the invariance of $p^*$, when introducing local contact  heterogeneities without varying the mobility patterns, implies that the spatial distribution of cases close to the epidemic threshold --controlled by the components of the  eigenvector of matrix ${\bf M}$-- is ruled by the structure of the underlying mobility network. We also observe that the value of the epidemic threshold at the peak $p^*$ is independent of the mobility network but is instead determined by the local heterogeneities, the difference in mixing of the subpopulations.}
	
	Finally, we extend our analysis to cover populations distributed heterogeneously across the metapopulation. In particular, we are interested in determining how the population asymmetry $\alpha$ and the local connectivity heterogeneity $\eta$ shape the relative magnitude of the peak of the epidemic threshold. To this aim, we represent  $\blbc (\alpha,\beta,\gamma;p^*)/\blb_0(\beta,\gamma)$ in figure~\ref{fig:phasediagram}, for $n_l = \alpha n_h$, $\av{k}_l = \beta \av{k}_h$, and $\av{k^2}_l = \gamma \av{k^2}_h$, in which $\gamma$ is given by equation~\eqref{eq:gamma} for the constraints imposed in section~\ref{sub:star}. We can observe that, as in figure~\ref{fig:threshold-curves}(b), increasing the local heterogeneity of the hub (lowering $\gamma$) increases the beneficial effect of the population mixing, as shown in figure~\ref{fig:phasediagram}(a). Interestingly, if we fix $\gamma$ and study the dependence of $\blbc (\alpha,\beta,\gamma;p^*)/\blb_0(\beta,\gamma)$ with $\alpha$ and $\beta$, as shown in figure~\ref{fig:phasediagram}(b), we observe that the detriment effect becomes stronger for larger values of $\beta$ since $\kmax$ increases so to keep $\gamma$ constant. In the opposite direction, when reducing the population of the periphery nodes, {\em i.e.}, decreasing $\alpha$, agents in the leaves are not able to substantially modify the connectivity distribution of residents in the hub, thus hindering the detriment effect in all investigated cases.
	
	\section{Conclusions}
	
	Driven by the advance of data mining techniques in mobility and social patterns~\cite{Guimera2005,Chowell2003,Patuelli2007}, epidemic models are continuously refined to bridge the gap existing between their theoretical predictions and the outcomes of real epidemic scenarios. In particular, within the very diverse realm of epidemic models, the proliferation of data sets capturing human movements across fine spatial scales have prompted the evolution of metapopulation frameworks, which constitute the usual approach to study the interplay between human mobility and disease spreading.  In this sense, the first theoretical frameworks assuming the population to move as random walkers across synthetic metapopulations~\cite{colizza2007reaction} have given rise to models incorporating the recurrent nature of human mobility~\cite{gmezgardees2017,granell2018epidemic,LFeng2020,belik2011}, the socio-economic facets of human movements~\cite{physrevx8031039,bosetti2020heterogeneity} or high-order mobility patterns~\cite{Matamalas16}.
	
	While most of the advances previously described have been focused on capturing the mobility flows more accurately, less attention has been paid to improve the contact patterns within each subpopulation. With few exceptions, such as the model recently proposed in \cite{Parino2021} incorporating the time varying nature of social contacts, human interactions are usually modeled using well-mixing hypothesis that do not capture the heterogeneous nature of human interactions and the role that this social heterogeneity has on the so-called super-spreading events. 
	
	In this work, we tackle this challenge and adapt the metapopulation model presented in reference~\cite{gmezgardees2017} to account for the heterogeneity in the number of contacts made by individuals. We describe a complete set of Markovian equations for a discrete-time Susceptible-Infected-Susceptible dynamics on subpopulations with recurrent mobility patterns. {These equations characterize the spatio-temporal evolution of the number of infected individuals across the system and show a good agreement with extensive agent-based simulations results. Computationally, iterating the equations of our formalism is orders of magnitude faster than performing the simulations because the latter should account for each microscopic stochastic process occurring in the population at each time step. Apart from the computational advantages, our formalism allows for deriving analytical results on the interplay between epidemics, mobility, and the structure of contacts within the metapopulation. Specifically, the linearization of these equations yields an accurate expression for the epidemic threshold, which is a crucial indicator for the design of interventions aimed at mitigating emerging outbreaks.}
	
	 {Our most important finding here is the emergence of the epidemic detriment when enhancing mobility, despite the fact that the individuals preserve their number of interactions independently of the visited locations. This result cannot be explained following the macroscopic arguments proposed in reference~\cite{gmezgardees2017} and shed light on the microscopic nature of the epidemic detriment phenomenon.} In particular, it becomes clear that this phenomenon is inherent to the variation of the contact structure of the population driven by redistribution of its individuals. Specifically, close to the epidemic threshold, the outbreak is mainly sustained by super-spreaders and the ties existing among them, which are weakened due to the homogenization of the underlying connectivity distributions caused by human mobility. Interestingly, the epidemic detriment observed in critical regimes is reversed in the super-critical regimes, where mobility increases epidemic prevalence, for it increases the average number of potentially infectious contacts made by scarcely connected individuals.
	
	The formalism here presented constitutes a step forward to account for the interplay between contact and flow structures and thus present several limitations. First of all, we assume that the number of interactions of each individual is constant and depends on the features of her residence patch, regardless of the place to which they move. Although this assumption can be interpreted as the preservation of the sociability of individuals, it prevents us from accounting for super-spreading events~\cite{althouse2020} associated to events or particular gatherings in which social connectivity is punctually amplified.  In addition, as remarked in the former paragraph, the results here obtained rely on assuming uncorrelated connectivity distributions within each patch. In this context, the effect of degree-degree correlations inside the patches deserves to be investigated; for example, one could expect the epidemic detriment to lose relevance in assortative populations, where ties connecting super-spreaders are strengthened and less likely to be influenced by the mobility. {Finally, although we have explored the physics of the interplay between contact heterogeneity and recurrent mobility with simple synthetic metapopulation networks, the model represents a general framework that can accommodate any arbitrary set of degree distributions within a population and any mobility network structure. In this sense, when data is available, the model can be investigated using a data-driven approach in the sense that one can easily include real data of demographics, mobility, and contact patterns to describe more realistic situations.}
	
	\ack
	
	W.C.\ acknowledges financial support from the Coordena\c{c}\~{a}o de Aperfei\c{c}oamento de Pessoal de N\'{i}vel Superior, Brazil -- CAPES, Finance Code 001. A.A. acknowledges financial support from Spanish MINECO (grant PGC2018-094754-BC21), Generalitat de Catalunya (grant No. 2017SGR-896 and 2020PANDE00098), and Universitat Rovira i Virgili (grant No. 2019PFR-URVB2-41). A.A. also acknowledges support from Generalitat de Catalunya ICREA Academia, and the James S. McDonnell Foundation (grant 220020325). 
	W.C. and S.C.F. acknowledge financial support from CAPES (grant No. 88887.507046/2020-00),
	{Conselho Nacional de Desenvolvimento Cient\'{i}fico e Tecnol\'{o}gico} -- CNPq
	(grants No. 430768/2018-4 and 311183/2019-0) and {Funda\c{c}\~{a}o de Amparo \`{a}
		Pesquisa do Estado de Minas Gerais} -- FAPEMIG (grant No. APQ-02393-18).
	D.S.P. and J.G.G. acknowledge financial support from MINECO (projects FIS2015-71582-C2 and FIS2017-87519-P), from the Departamento de Industria e Innovaci\'on del Gobierno de Arag\'on y Fondo Social Europeo (FENOL group E-19), and from Fundaci\'on Ibercaja and Universidad de Zaragoza (grant 224220). 
	
	\appendix
	
	\section{Exact evaluation of the epidemic threshold for a star-like metapopulation}
	\label{app:new-exactStar}
	
	In this case, we have to evaluate seven different terms:
	\begin{itemize}
		\item $M_{hh}$: contact of two individuals residing in the hub;
		\item $M_{lh}$: contact of one resident from a leaf with another from the hub;
		\item $M_{hl}$: contact of one resident from the hub with another from a leaf;
		\item $M_{ll}$: contact of two individuals residing in the same leaf;
		\item $M_{l,l+1}$: contact of one resident from a leaf with another from its adjacent leaf;
		\item $M_{l,l-1}$: contact of one resident from the adjacent leaf with one from the other leaf;
		\item $M_{ln}$: contact of two residents from different and not adjacent leaves;
	\end{itemize}
	
	The mobility matrix elements $R_{ij}$ are expressed in \cref{eq:starRhl,eq:starRlh,eq:starRllp1}. Applying these expressions in \eqref{eq:new_matrizMind}, we have
	
	\numparts
	\begin{eqnarray}
		\label{eq:hubMterms}
		M_{hh} &= \av{k^2}_h \left[ (1-p)^2\ffrac{1}{Q_h} + \ffrac{p^2}{\kappa}\ffrac{1}{Q_l}\right]n_h,\\
		M_{lh} &= \av{k^2}_h \left[   (1-p)p\left(    \ffrac{1}{\kappa}\ffrac{1}{Q_l} + \delta\ffrac{1}{Q_h}  \right) + p^2 \ffrac{(1-\delta)}{\kappa} \ffrac{1}{Q_l} \right]n_h, \\
		M_{hl} &= \av{k^2}_l \left[  (1-p)p\left(  \delta \ffrac{1}{Q_h} + \ffrac{1}{\kappa} \ffrac{1}{Q_l}\right) + p^2 \ffrac{(1-\delta)}{\kappa} \ffrac{1}{Q_l}  \right]n_l, \\
		M_{ll} &= \av{k^2}_l \left[   (1-p)^2\ffrac{1}{Q_l} + p^2 \ffrac{(1-\delta)^2}{Q_l} + p^2\ffrac{\delta^2}{Q_h}     \right] n_l, \\
		M_{l,l+1} &= \av{k^2}_l \left[    (1-p)p\ffrac{(1-\delta)}{Q_l} + p^2 \ffrac{\delta^2}{Q_h}\right]n_l,  \\
		M_{l,l-1} &= \av{k^2}_l \left[    (1-p)p\ffrac{(1-\delta)}{Q_l} + p^2 \ffrac{\delta^2}{Q_h}\right]n_l,  \\ 
		M_{ln} &= \av{k^2}_l \left( p^2\ffrac{\delta^2}{Q_h}\right)n_l.
	\end{eqnarray}\endnumparts
	
	Again, by evaluating equation~\eqref{eq:new_eigeneqind}, we have, for the hub,
	\begin{eqnarray}
		\bmu \bepik{h} = \blb \sum_j M_{hj} \bepik{j}=  \blb M_{hh} \bepik{h} + \kappa \blb M_{hl} \bepik{l},
	\end{eqnarray}
	while for a leaf we have
	\begin{eqnarray}
		\bmu \bepik{l} = \blb \sum_j M_{lj}\bepik{j} &=&  
		\blb M_{lh} \bepik{h} + 
		\blb M_{ll} \bepik{l} \nonumber \\
		&+& 
		\blb M_{l,l+1} \bepik{l} + 
		\blb M_{l,l-1} \bepik{l} + 
		\blb (\kappa - 3) M_{ln} \bepik{l},
	\end{eqnarray}
	in which the factor $3$ in the last term is since there are $\kappa -3$ other leafs not directly connected to a single leaf ($R_{ln} = 0$). The statistical equivalence of the leaves allows us to recast the computation of the epidemic threshold in a eigenvalue problem of a $2\times 2$ matrix
	\begin{equation}
		{\bf M} = 
		\begin{pmatrix}
			M_{hh} &~~~ \kappa M_{hl} \\
			M_{lh} &~~~ M_{ll} + M_{l,l+1} + M_{l,l-1} + (\kappa-3)M_{ln},
		\end{pmatrix}.
	\end{equation}
	The leading eigenvalue will be given by $\Lambda_\mathrm{max} = \ffrac{\Tr M + \sqrt{(\Tr M)^2 - 4 \det M}}{2}$, that was solved using SymPy~\cite{sympy} to get the results shown in the main text.
	
	\section{Epidemic threshold in the static case}
	\label{app:new-static}
	
	To check the consistency of these equations, let us consider the static case in which all individuals stay in their patches and do not move: $p_k = 0~\forall k$. So, equation~\eqref{eq:new_matrizMind} becomes
	\[
	\left.\Mij\right|_{p_k = 0} = \av{k^2}_j \ffrac{\dij}{\left.Q_i\right|_{p_k=0}} n_j,
	\]
	where
	$
	\left.Q_i\right|_{p_k=0} = n_i \av{k}_i,
	$
	that after being used in \eqref{eq:new_eigeneqind} results in
	$
	\bmu \bepik{i} = \blb \ffrac{\av{k^2}_i}{\av{k}_i}    \bepik{i}.
	$
	This case consists of isolated subpopulations in an annealed regime in which the epidemic threshold will be given by the first subpopulation in the active state, if its population is not so small compared to other patches. Indeed, the usual epidemic threshold known in the HMF theory is obtained, 
	\begin{equation}
		\blbc = \bmu\min_i\left\lbrace\ffrac{\av{k}_i}{\av{k^2}_i}\right\rbrace.
		\label{eq:new-staticblb}
	\end{equation}
	Therefore, in the static case the epidemic threshold of the metapopulation corresponds to the individual epidemic threshold of the most vulnerable patch. 
	
	\section{Perturbative analysis of the epidemic threshold}
	\label{app:new-firstSecond}
	
	We proceed by making a perturbative analysis of the eigenvalues of the matrix {$\bf M$} up to first order on $p$ to complement the discussions of the main text. First, it is convenient to rewrite equation~\eqref{eq:new_matrizMind} to split the terms with different order in $p$:
	\begin{eqnarray}
		\Mij = \av{k^2}_j \left\{     \ffrac{\dij}{Q_i} + p\left[  
		\ffrac{R_{ji}}{Q_i} + \ffrac{R_{ij}}{Q_j} -2\ffrac{\dij}{Q_i}
		\right] 
		+
		\right.	\nonumber\\\left.\qquad\qquad\qquad\qquad
		p^2 \left[
		\ffrac{\dij}{Q_i} - \ffrac{R_{ji}}{Q_i} - \ffrac{R_{ij}}{Q_j} +
		\sum_l \ffrac{R_{il}R_{jl}}{Q_l}
		\right]		
		\right\}n_j.
		\label{eq:Mij-pert1}
	\end{eqnarray}
	Since $Q_i$ is also a function of $p$, we must perform a Taylor expansion around $p= 0$, knowing that $\left.Q_i\right|_{p=0} = \nkm{i}$. 
	The first derivative of $Q_i$ is
	\[
	\left.\derivative{Q_i}{p}\right|_{p=0} = \sum_j \av{k}_j \left( R_{ji} - \delta_{ij} \right) n_j.
	\]
	Let us define
	\[
	r_i \equiv \sum_j  \left(  - R_{ji} + \dij \right) n_j \av{k}_j,
	\]
	so that
	\[
	\left.\derivative{}{p}\left(\ffrac{1}{Q_i}\right)\right|_{p=0}  = \ffrac{r_i}{\left(n_i \av{k}_i\right)^2}.
	\]
	Next,  keeping only terms up to order 1, we have
	\[
	\ffrac{1}{Q_i} = \ffrac{1}{\nkm{i}} + p \ffrac{r_i}{\left(n_i \av{k}_i\right)^2} + \mathcal{O}(p^2).
	\]
	Substituting the last expression in \eqref{eq:Mij-pert1} we get, after some algebra,
	\begin{equation}
		\label{eq:new-pertbM}
		\Mij = \tMij{0} + p\tMij{1} + \mathcal{O}(p^2), 
	\end{equation}
	where
	\numparts\begin{eqnarray}
		\label{eq:new-pertbM_term}
		\label{eq:new-pertbM_term0}
		\tMij{0}  = \dij  \ffrac{\av{k^2}_i}{\av{k}_i},
		\\\nonumber
		\\
		\label{eq:new-pertbM_term1}
		\tMij{1}  = \left[ 
		\ffrac{R_{ij}}{\nkm{j}} + \ffrac{R_{ji}}{\nkm{i}} + \ffrac{\dij}{\nkm{i}} \left(    \ffrac{r_i}{\nkm{i}} - 2   \right)
		\right] n_j \av{k^2}_j.
	\end{eqnarray}\endnumparts
	
	From the static case, we know that there are $\Omega$ unperturbed eigenvalues $\Lb{0}_i = \av{k^2}_i/\av{k}_i$, for $p = 0$, with normalized eigenvectors $\vec{\bepik{}}_i = \left\{ \bepik{j} \right\}$ and $\bepik{j} = \dij$; see equation~\eqref{eq:new-staticblb}. Assuming that the eigenvalues are not degenerate, the new eigenvalues will be given by~\cite{marcus2001brief}
	\begin{equation}
		\Lambda_i \approx \Lb{0}_i + p \Lb{1}_i,
		\label{eq:lambdaantes}
	\end{equation}
	\begin{figure}[t!]
		\centering
		\includegraphics[width=\linewidth]{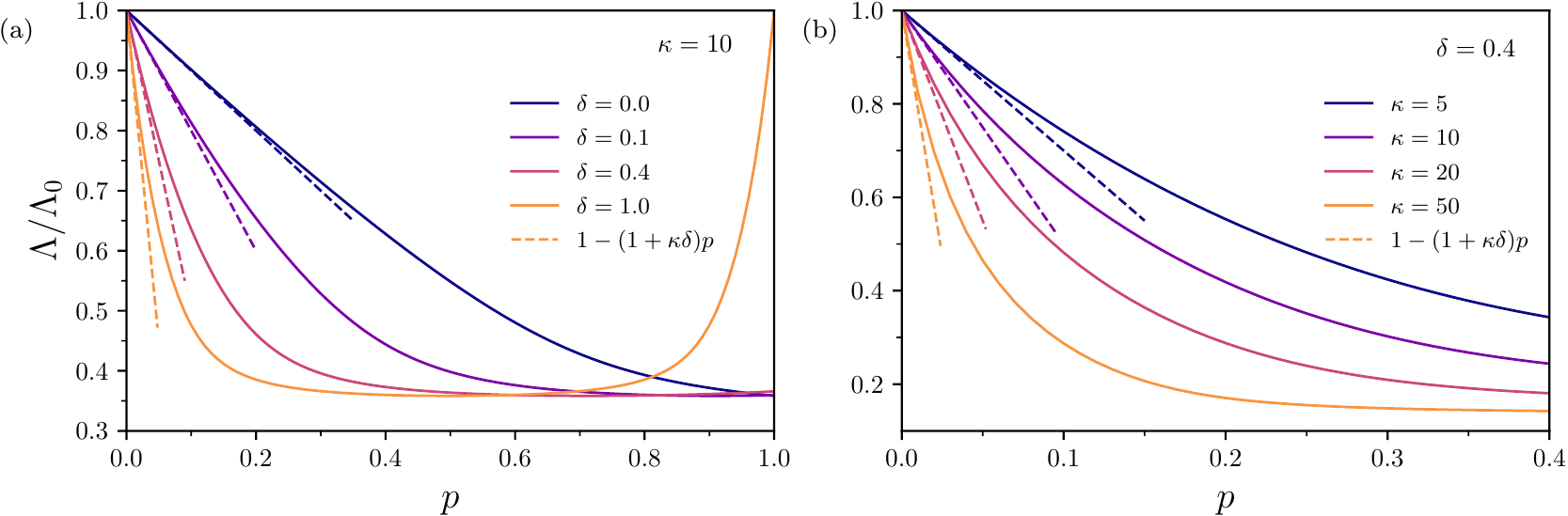}
		\caption{Normalized leading eigenvalue of matrix ${\bf M}$ as a function of the mobility for different values of the number of leaves $\kappa$ and the agents flow from leaves to the hub governed by $\delta$, with the same number of individuals ($\alpha = 1$). Solid lines show the exact values whereas dotted lines corresponds to the linear correction estimated by the perturbative approach via equation~\eqref{eq:pertlinearcorrection}. We fix the number of leaves $\kappa$ and modify $\delta$ (color code) in (a) and present the complementary analysis in (b).
		}
		\label{fig:pertubcompara}
	\end{figure}
	where
	\numparts
	\begin{eqnarray}
		\label{eq:new-pertubLbs}
		\label{eq:new-pertubLbs0}
		\Lb{0}_i = \ffrac{\av{k^2}_i}{\av{k}_i},
		\\\nonumber \\
		\label{eq:new-pertubLbs1}
		\Lb{1}_i = \vec{\bepik{}}_i \tM{1} \vec{\bepik{}}_i.
	\end{eqnarray}\endnumparts
	
	Substituting equation~\eqref{eq:new-pertbM_term1} in \eqref{eq:new-pertubLbs1}, after some algebra we get the first correction to the eigenvalue,
	\begin{equation}
		\label{eq:pertlinearcorrection}
		\ffrac{\Lb{1}_i}{\Lb{0}_i} =  
		R_{ii} - 1 - \sum_{j\neq i} R_{ji} \ffrac{\nkm{j}}{\nkm{i}}
		.
	\end{equation}
	
	Interestingly, unlike the original MIR model, the first order correction depends on the underlying topology. To check the accuracy of this correction, we represent in figure~\ref{fig:pertubcompara} the leading eigenvalues of the matrix ${\bf M}$ along with the linear correction provided by the perturbative analysis, finding a remarkable agreement in the low mobility regime $p\ll1$.
	
	\section*{References}
	\bibliography{refs.bib}
	\bibliographystyle{iopart-num}
	
\end{document}